\documentclass[sigconf,nonacm]{acmart}

\AtBeginDocument{%
  \providecommand\BibTeX{{%
    \normalfont B\kern-0.5em{\scshape i\kern-0.25em b}\kern-0.8em\TeX}}}

\begin{document}

\title{Pynblint: a Static Analyzer for Python Jupyter Notebooks}

\author{Luigi Quaranta}
\affiliation{%
  \institution{University of Bari}
  \streetaddress{Via Edoardo Orabona, 4}
  \city{Bari}
  \country{Italy}
  \postcode{70125}
}
\email{luigi.quaranta@uniba.it}
\orcid{0000-0002-9221-0739}

\author{Fabio Calefato}
\affiliation{%
  \institution{University of Bari}
  \streetaddress{Via Edoardo Orabona, 4}
  \city{Bari}
  \postcode{70125}
  \country{Italy}}
\email{fabio.calefato@uniba.it}
\orcid{0000-0003-2654-1588}

\author{Filippo Lanubile}
\affiliation{%
  \institution{University of Bari}
  \streetaddress{Via Edoardo Orabona, 4}
  \city{Bari}
  \postcode{70125}
  \country{Italy}}
\email{filippo.lanubile@uniba.it}
\orcid{0000-0003-3373-7589}

\begin{abstract}
Jupyter Notebook is the tool of choice of many data scientists in the early stages of ML workflows.
The notebook format, however, has been criticized for inducing bad programming practices;
indeed, researchers have already shown that open-source repositories are inundated by poor-quality notebooks.
Low-quality output from the prototypical stages of ML workflows constitutes a clear bottleneck towards the productization of ML models.
To foster the creation of better notebooks, we developed Pynblint, a static analyzer for Jupyter notebooks written in Python. The tool checks the compliance of notebooks (and surrounding repositories) with a set of empirically validated best practices and provides targeted recommendations when violations are detected.
\end{abstract}

\keywords{Computational notebooks, lint, software quality, data science, machine learning}

\begin{teaserfigure}
  \vspace{0.8cm}
\end{teaserfigure}

\maketitle

\section{Introduction}

The massive adoption of AI-based technologies in the modern software industry is raising new intriguing challenges, many of which concern the shift of machine learning (ML) model prototypes into production-ready software components.
Often, a variety of factors contribute to rendering this shift difficult and costly to achieve; these range from technical matters -- like the complexity of reproducing lab model performances in live systems -- to human aspects, arising from the coexistence of varied backgrounds and perspectives in multidisciplinary teams \cite{nahar_collaboration_2021}.

In this context, a prominent role is played by the tools that practitioners adopt at the various stages of ML workflows. Tool misuse can hinder the quality of team collaboration and constitute a bottleneck towards the productization of ML models.
Notably, computational notebooks represent a prime example. In the last few years, they have established themselves as the tool of choice of many data scientists for activities comprised in the early stages of ML workflows, from data exploration to model prototyping. Their most popular implementation, Jupyter Notebook,\footnote{https://jupyter.org} combines code, documentation, and multimedia output in an interactive narrative of computations, providing unparalleled support for fast experimental iterations and lightweight documentation of experiments.
However, besides the evident benefits they bring, Jupyter notebooks have been criticized for inducing bad programming habits and offering limited built-in support for software engineering best practices \cite{chattopadhyay2020s, joel_grus_i_2018}. Indeed, researchers have already shown that notebooks often contain poor quality code and that -- in spite of their original vocation -- they typically end up being messy and scarcely documented \cite{wang_better_2020, pimentel_understanding_2021}.

Commonly, poor-quality and hardly-reproducible notebooks, written by data scientists in the early stages of ML workflows, get in the way of the model productization process. Indeed, transitioning from ML model prototypes to production-ready ML components often entails, in practice, the consolidation of experimental code from notebooks into structured and tested codebases \cite{lanubile_towards_2021}. Under such circumstances, low-grade notebooks might represent an expensive bottleneck and a potential source of technical debt.

\section{Pynblint}

In our previous work \cite{quaranta_eliciting_2022}, we collected and validated a catalog of 17 best practices for professional collaboration with computational notebooks. Our guidelines foster a use of notebooks aware of software engineering best practices, with the aim of boosting their benefits while preventing potential drawbacks. In the light of these findings, we have developed Pynblint,\footnote{https://github.com/collab-uniba/pynblint} a static analysis tool for Jupyter notebooks written in Python.

Besides being the most popular notebook platform to date, Jupyter Notebook has inspired the design of most modern computational notebook implementations; the majority of them currently adopts the \texttt{.ipynb} JSON-encoded format and offers the same core functionalities as Jupyter (e.g., Google Colaboratory\footnote{https://colab.research.google.com}). Furthermore, most Jupyter notebooks are written in Python. For these reasons we chose Python Jupyter notebooks as the target of our static analysis tool.

\subsection{Definition of the linting rules}

Pynblint implements a set of checks to assess the quality of notebooks and the code repositories they belong to. We derived each check (hereafter referred to as \textit{linting rule}) as an operationalization of the best practices from our catalog. These best practices range from recommendations for better traceability and reproducibility of computations (e.g., \textit{``Use version control''} and \textit{``Manage project dependencies''}) to hints for code quality enhancements (e.g., \textit{``Stick to coding standards''} and \textit{``Test your code''}), and clues for a more consistent use of notebooks narrative capabilities (e.g., \textit{``Leverage Markdown headings to structure your notebook''}).

Not all best practices could be fully operationalized. For instance, we found no practical way to verify compliance with the recommendation \textit{``Distinguish production and development artifacts''}.
In other cases, we resorted to partial operationalizations. For example, compliance with the best practice \textit{``Make your data available''} is verified by detecting the use of DVC, a git-based Data Version Control system with support for cloud remotes. We plan to extend these operationalizations in the near future by considering larger sets of implementations. Meanwhile, the related linting rules can be disabled if they do not apply to specific professional settings.

In general, customization has been a primary concern in the design of Pynblint. Not only the predefined linting rules can be dynamically included or excluded from the analysis, but the linting engine itself features a plug-in architecture that enables Pynblint users to add their own linting rules to the core set.

\subsection{Interface}

Pynblint offers a command-line interface (CLI) capable of displaying detailed linting results, including the preview of flawed cells.
The tool accepts three main types of input: (1) standalone Python Jupyter Notebooks, to be analyzed in isolation; (2) local code repositories containing Jupyter notebooks (both in the form of uncompressed directories or compressed \texttt{.zip} archives); (3) GitHub public repositories containing Jupyter notebooks. In the future, we will extend the array of available input types by including, for instance, standalone Google Colab notebooks as well as private GitHub repositories.
Other than rendered in the terminal, results from the linting process can be saved as Markdown-formatted reports or serialized in a machine-readable \texttt{JSON} format, allowing further post-processing. Exporting results in the \texttt{HTML} format is on our roadmap.

\subsection{Usage}

Pynblint is available on PyPI, the official Python Package Index, and therefore can be installed via package managers such as \texttt{pip} and \texttt{poetry}.
Once available in the active Python environment, the tool can be used to analyze standalone notebooks (e.g., \texttt{pynblint Example.ipynb}) as well as code repositories containing notebooks (e.g., \texttt{pynblint example/project/path}).

When analyzing the working directory (i.e., \texttt{pynblint .}), the linter will start by looking at the contents of the project root. Then, it will recursively seek Python Jupyter notebooks in all existing sub-directories. At the end of the process, the results are rendered in the terminal. 

Additional options can be tweaked for customized behavior; for instance, to save linting results in a Markdown report, an output filename with the \texttt{.md} extension should be specified (e.g., \texttt{--output report.md}).
More conveniently, recurring options can be declared in a dotenv file named \texttt{.pynblint}, to be placed in the folder from which the linter is invoked (typically, the project root).

As a static analyzer, Pynblint can be also leveraged in the context of CI/CD pipelines. For instance, once installed on a CI/CD server (e.g., GitHub Actions), it can be invoked alongside other quality assurance tools at build time. According to user preferences, a build might fail if the linting process reveals potential notebook defects. 

This feature sets Pynblint apart from Julynter~\cite{pimentel_understanding_2021}, which is -- to the best of our knowledge -- the only alternative Jupyter linting tool. Julynter was developed as a plug-in of Jupyter Lab, the evolution of the Jupyter Notebook IDE; it performs real-time checks on the quality and reproducibility of Jupyter notebooks while also providing improvement recommendations. However, Julynter can only be executed as a live assistant within Jupyter Lab and cannot be leveraged off-line, as a standalone module. Therefore, while useful during notebook writing, Julynter cannot be integrated into CI/CD pipelines and cannot be adopted as a pre-commit hook. 
Moreover, being tied to the Jupyter Lab environment, it cannot be used to analyze notebooks produced with different platforms (e.g., Google Colab or the Kaggle Notebooks IDE), even if they comply with the standard Jupyter format (\texttt{.ipynb}).

\section{Conclusion and Future Work}

We implemented Pynblint, a static analyzer for Python Jupyter notebooks. To refine the tool, we are currently conducting a formative study with experienced Jupyter users; at the same time, we are developing a web front-end for Pynblint, to make it easily accessible by novice data scientists with limited or no command-line experience.
Furthermore, we will validate the tool with a field study, involving data science professionals from multiple companies.
As a result of the validation process, we expect to expand the core set of available linting rules and to make the tool proactive, i.e., capable of automatically fixing a selected set of violations.

\bibliographystyle{ACM-Reference-Format}
\bibliography{references}

\end{document}